\documentclass[letterpaper,11pt]{article}
\usepackage{fullpage}

\usepackage[]{newtxtext}
\usepackage[bigdelims,vvarbb]{newtxmath} 
\usepackage{url}
\usepackage{graphicx}
\usepackage[natbibapa]{apacite}
\usepackage{subcaption}

\def\nA{n_\mathrm{Control}}
\def\nB{n_\mathrm{AUI}}

\title{Autocompletion interfaces make crowd workers slower, but their use promotes response diversity}
\author{Xipei Liu\affil{University of Vermont}
        \and James P.~Bagrow\affil{University of Vermont}}

\author{Xipei Liu \qquad James P.~Bagrow\\
{\small University of Vermont,}\\
{\small Vermont Complex Systems Center,}\\
{\small Burlington, VT, USA}}

\begin{document}

\maketitle

\begin{abstract}
Creative tasks such as ideation or question proposal are powerful applications of crowdsourcing, yet the quantity of workers available for addressing practical problems is often insufficient.
To enable scalable crowdsourcing thus requires gaining all possible efficiency and information from available workers.
One option for text-focused tasks is to allow assistive technology, such as an autocompletion user interface (AUI), to help workers input text responses. 
But support for the efficacy of AUIs is mixed.
Here we designed and conducted a randomized experiment where workers were asked to provide short text responses to given questions.
Our experimental goal was to determine if an AUI helps workers respond more quickly and with improved consistency by mitigating typos and misspellings.
Surprisingly, we found that neither occurred: workers assigned to the AUI treatment were slower than those assigned to the non-AUI control and their responses were more diverse, not less, than those of the control.
Both the lexical and semantic diversities of responses were higher, with the latter measured using word2vec.
A crowdsourcer interested in worker speed may want to avoid using an AUI, but using an AUI to boost response diversity may be valuable to crowdsourcers interested in receiving as much novel information from workers as possible.
\end{abstract}

\section{Introduction}

Crowdsourcing applications vary from basic, self-contained tasks such as image recognition or labeling~\citep{welinder2010online} all the way to open-ended and creative endeavors such as collaborative writing, creative question proposal, or more general ideation~\citep{little2010exploring}.
Yet scaling the crowd to very large sets of creative tasks may require prohibitive numbers of workers.
Scalability is one of the key challenges in crowdsourcing: how to best apply the valuable but limited resources provided by crowd workers and how to help workers be as efficient as possible. 

Efficiency gains can be achieved either collectively at the level of the entire crowd or by helping individual workers.
At the crowd level, efficiency can be gained by assigning tasks to workers in the best order~\citep{tran2013efficient}, by filtering out poor tasks or workers, or by best incentivizing workers~\citep{allahbakhsh2013quality}.
At the individual worker level, efficiency gains can come from helping workers craft more accurate responses and  complete tasks in less time.

One way to make workers individually more efficient is to computationally augment their task interface with useful information.
For example, an autocompletion user interface (AUI)~\citep{Sevenster2012107},  such as used on Google's main search page, may speed up workers as they answer questions or propose ideas.
However, support for the benefits of AUIs is mixed and existing research has not considered short, repetitive inputs such as those required by many large-scale crowdsourcing problems.
More generally,
it is not yet clear what are the best approaches or general strategies to achieve efficiency gains for creative crowdsourcing tasks.

In this work, we conducted a randomized trial of the benefits of allowing workers to answer a text-based question with the help of an autocompletion user interface. 
Workers interacted with a web form that recorded how quickly they entered text into the response field and how quickly they submitted their responses after typing is completed. 
After the experiment concluded, we measured response diversity using textual analyses and response quality using a followup crowdsourcing task with an independent population of workers.
Our results indicate that the AUI treatment did not affect quality, and did not help workers perform more quickly or achieve greater response consensus. Instead, workers with the AUI were significantly slower and their responses were more diverse than workers in the non-AUI control group.

\section{Related Work}
\label{sec:relatedwork}

An important goal of crowdsourcing research is achieving
efficient scalability of the crowd to very large sets of tasks.
Efficiency in crowdsourcing manifests both in receiving more effective information per worker and in making individual workers faster and/or more accurate.
The former problem is a significant area of interest~\citep{doi:10.1287/opre.2013.1235,li2016crowdsourcing,mcandrew2016reply} while less work has been put towards the latter.

One approach to helping workers be faster at individual tasks is the application of usability studies.
\citeauthor{kittur2008crowdsourcing} (\citeyear{kittur2008crowdsourcing}) famously showed how crowd workers can perform user studies, although this work was focused on using workers as usability testers for other platforms, not on studying crowdsourcing interfaces.
More recent usability studies on the efficiency and accuracy of workers include: \citeauthor{cheng_break_2015} (\citeyear{cheng_break_2015}), who consider the task completion times of macrotasks and microtasks and find workers given smaller microtasks were slower but achieve higher quality than those given larger macrotasks; \citeauthor{lasecki_effects_2015} (\citeyear{lasecki_effects_2015}), who study how the sequence of tasks given to workers and interruptions between tasks may slow workers down; and
\citeauthor{demartini2016crowdsourcing} (\citeyear{demartini2016crowdsourcing}), who study completion times for relevance judgment tasks, and find that imposed time limits can improve relevance quality, but do not focus on ways to speed up workers.
These studies do not test the effects of the task interface, however, as we do here.

The usability feature we study here is an autocompletion user interface (AUI). AUIs are broadly familiar to online workers at this point, thanks in particular to their prominence on Google's main search bar (evolving out of the original Google Instant implementation).
However, literature on the benefits of AUIs (and related word prediction and completion interfaces) in terms of improving efficiency is decidedly mixed.

It is generally assumed that AUIs make users faster by saving keystrokes~\citep{bast2006type}.
However, there is considerable debate about whether or not such gains are countered by increased cognitive load induced by processing the given autocompletions~\citep{331567}. 
\citeauthor{anson2006effects}
(\citeyear{anson2006effects}) showed that typists can enter text more quickly with word completion and prediction interfaces than without. However, this study focused on a different input modality (an onscreen keyboard) and, more importantly, on a text transcription task: typists were asked to reproduce an existing text, not answer questions.
\citeauthor{Sevenster2012107} (\citeyear{Sevenster2012107})
 showed that medical typists saved keystrokes when using an autocompletion interface to input standardized medical terms. However, they did not consider the elapsed times required by these users, instead focusing on response times of the AUI suggestions, and so it is unclear if the users were actually faster with the AUI.
There is some evidence that long-term use of an AUI can lead to improved speed and not just keystroke savings~\citep{magnuson2002measuring}, but it is not clear how general such learning may be, and whether or not it is relevant to short-duration crowdsourcing tasks.

\section{Experimental design}
\label{sec:experimentdesign}

Here we describe the task we studied and its input data, worker recruitment, the design of our experimental treatment and control,  the ``instrumentation'' we used to measure the speeds of workers as they performed our task, and our procedures to post-process and rate the worker responses to our task prior to subsequent analysis.

\begin{figure*}
    \centering
    \begin{subfigure}[b]{0.48\textwidth}  \frame{\includegraphics[width=\textwidth,trim=0 0 0 11,clip=true]{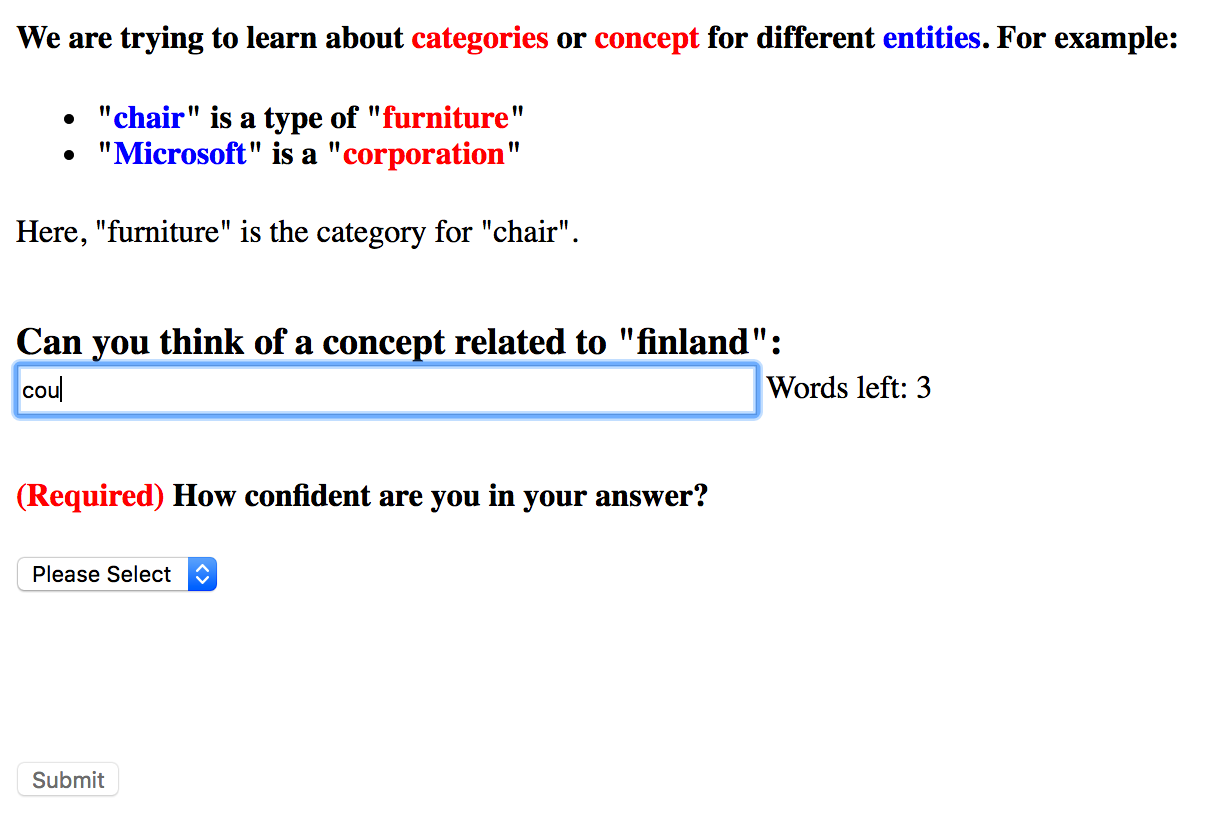}}
        \caption{Control form}
        \label{fig:screenshotA}
    \end{subfigure}
    \begin{subfigure}[b]{0.48\textwidth}        \frame{\includegraphics[width=\textwidth,trim=0 -5 0 0,clip=true]{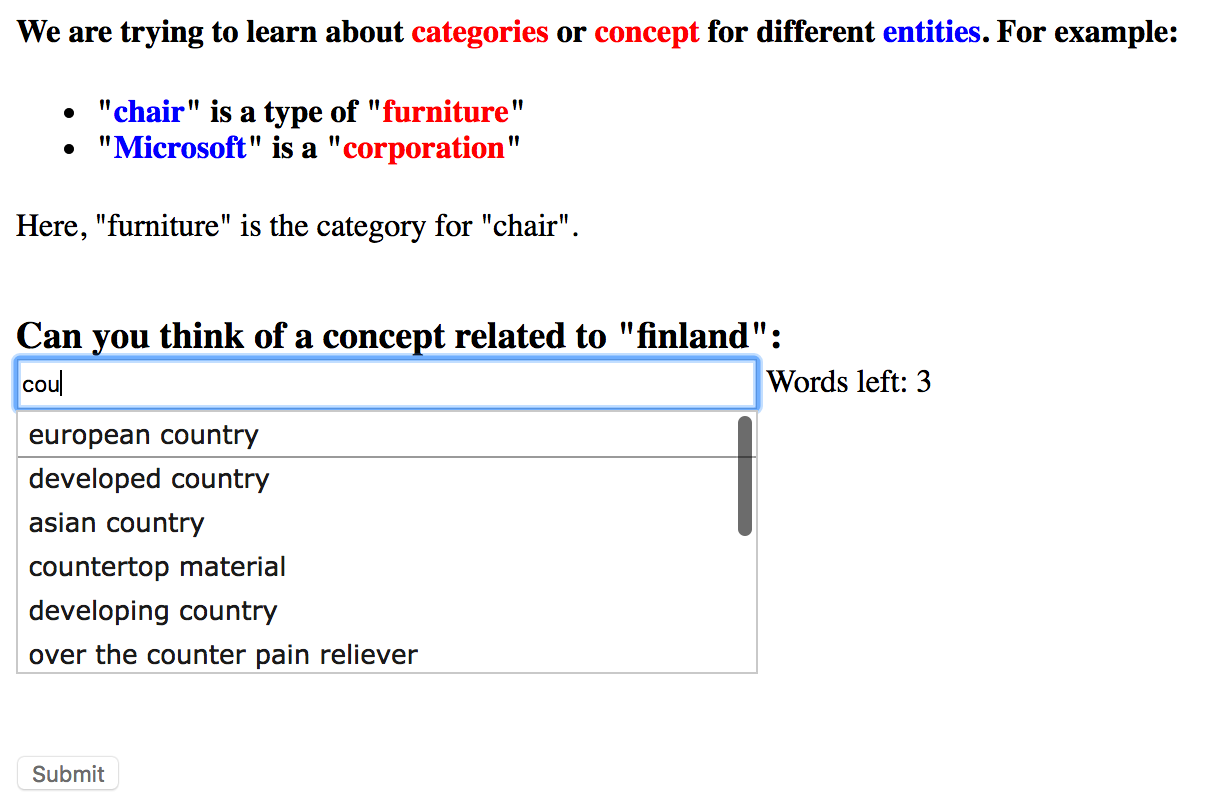}}
        \caption{Autocompletion User Interface (AUI) form}
        \label{fig:screenshotB}
    \end{subfigure}
    \caption{Screenshots of our conceptualization task interface. The presence of the AUI is the only difference between the task interfaces.
    \label{fig:screenshot}}
\end{figure*}

\paragraph{Task description and question data}
For this work, we focused on a conceptualization or ``IsA'' task. 
Each task consisted of a question of the form: ``\textit{FOO} is a type of:'' followed by a short one-line text field for the worker to respond.
The particular terms ``\textit{FOO}'' then defines each question.
Before this question was a brief description of the task followed by two examples: ``\textit{chair} is a type of furniture'' and ``\textit{Microsoft} is a corporation''. 
See Fig.~\ref{fig:screenshot}.

The question terms (``chair'' and ``Microsoft'' in the above examples) were
chosen from the Microsoft Concept Graph (MCG) dataset~\citep{wu2012probase,wang2015inference}. These data provide a bipartite knowledge graph linking \textit{entities} to \textit{concepts}, for example ``city'' is a concept related to the entity ``Berlin''. 
We chose these data for our conceptualization task so that we have a comparative baseline, as the MCG captures the same relationships we measure in our task.

We chose 10 entities randomly from the MCG to act as question terms.
The MCG data are somewhat noisy, heavily skewed to rare terms (often medical terms), and contain many abstract entity--concept relations, so we first performed a filtering step to focus on commonplace and easy-to-understand question terms.
We also required that 5 of the chosen terms be one-word entities longer than two letters and 5 be multi-word phrases, both without numbers.
See Table~\ref{tab:questions} for our final chosen question terms.

\begin{table}
\centering\small
\begin{tabular}{cccc}
\hline 
\emph{ID} & \emph{Question term}  & \emph{ID} & \emph{Question term} \\ 
\hline
Q1 & hail       & Q6  & occupational therapist\\
Q2 & millet   & Q7  & standard deviation\\
Q3 & steam  & Q8 & motor vehicle \\
Q4 & finland  & Q9 & dengue fever\\
Q5 & spider	  & Q10 & citric acid \\ 
\hline
\end{tabular}
\caption{Question terms used in our conceptualization task. 
Workers were shown these questions in random order.
\label{tab:questions}}
\end{table}

\paragraph{Crowdsourcing and treatment}

We recruited workers on Amazon Mechanical Turk (AMT) to perform our task.
Recruited workers must have 80\% or better approval rating, be USA-located, and be able to view adult content.
Each human intelligence task (HIT) was one conceptualization task, i.e. one of the ten questions. 
Workers could perform anywhere from one to ten HITs.
Questions were shown to each worker in random order.
Each worker response generates a question-response text pair which may or may not be unique as other workers may give the same response to the same question.
Workers were compensated \$0.05 per HIT.

Workers were blindly assigned to one of two conditions with equal probability (simple random assignment) when they accepted their first HIT.
This assignment was then carried over for any subsequent HITs performed by that worker.
The control group consisted of a HIT interface (web form) with a text entry field without an autocompletion user interface (AUI).
We refer to this as the Control form and the workers assigned to the Control form as the Control group.
The treatment consisted of a text entry field but with an associated AUI; corresponding to the Control group, we refer to this form as the AUI form and the workers assigned to the AUI form as the AUI group.
Screenshots comparing Control and AUI forms are shown in Fig.~\ref{fig:screenshot}.

In all other respects the HIT interfaces were identical. 
In particular, for both forms,
JavaScript was used on the field to prevent workers from inputting punctuation or responses exceeding four words.
Copy or paste is prevented on the page; workers can only fill in the text entry by typing or, if it is available, by selecting from the AUI.
The HIT was not submittable until the response field was filled.

\paragraph{Autocompletion user interface}

The AUI we used was implemented with jQuery-UI's (ver.\ 1.12.1) autocomplete widget with autofocus enabled\footnote{Autofocus makes it easy for the worker to quickly select the top AUI response.}.
Whenever two or more characters are present in the response field, a search based on the current contents of the response field is triggered of a database containing all MCG concepts with at least 5 associated entities ($n=$ 705,710 concept terms). 
Concept terms are indexed for speed and the search term is matched from both sides using MySQL's ``LIKE'' operator, and the first six matches are dynamically displayed in the AUI (Fig.~\ref{fig:screenshotB}) with up to another six available by scrolling.
The search repeats whenever the current response changes; the AUI disappears if there are fewer than two characters present in the response field.
Workers were not required to select a response from the AUI. 
Searching the MCG concepts helps provide meaningful autocompletions for our conceptualization task.

\paragraph{Instrumentation}

Our experimental goal was to determine how workers would use an AUI and how an AUI may affect their responses. Would they be faster at answering such short questions by saving on typing time? Or would the cognitive load of reading the AUI as it appeared and updated slow down the worker, even enough to offset any savings from faster text entry? Would the AUI lead to more consistent responses across workers by mitigating typos, or less consistent responses, by acting as a cognitive primer?

To study the effects of the AUI, each HIT form was instrumented with JavaScript
to record the times when workers first entered text into the response field, when they last entered text into the response field, and when the form was submitted.
Note that while we also recorded the time when the HIT was accepted, we did not use these data because it is unclear when a worker accepts a HIT as opposed to when a worker actually begins work on that HIT (AMT workers sometimes open a series of HITs into separate browser tabs, and then later process those HITs). 
Due to this, our future experiments will also record when the browser window containing the HIT is active.

This instrumentation allows us to measure two important features of worker activity:
\begin{enumerate}
\item Typing duration---Total elapsed time between the first and last keypress made by the worker into the text area.
\item Submission delay---Total elapsed time between the final keypress into the text area and the submission of the form.
\end{enumerate}

\paragraph{Response processing and quality ratings}
Worker responses were post-processed by removing casing and transforming any whitespace to a single space character. Additional processing was unnecessary because of the in-browser processing done by the form (see above).

A second, non-experimental set of HITs was used to measure the perceived quality of each unique question-response pair. 
Instead of using additional workers to rate responses, the
quality of responses for our conceptualization task could be assessed computationally using, for example, ontology datasets. However, combining free text responses from workers with a fixed-vocabulary dataset is a challenging natural language processing task beyond the scope of this work, so here we simply relied on ratings by independent workers.
Workers were shown statements of the form ``\textit{FOO} is a type of: \textit{BAR}'', where \textit{BAR} is a worker response to question term \textit{FOO}, and asked to rate their agreement with this statement on a 1--5 rating scale (1---least agree; 5---strongest agree).
Each worker was shown ten such statements per HIT, and compensated at a rate of \$0.25 per HIT.
Workers who belong to either Control or AUI groups were excluded from these tasks.

\section{Results}
\label{sec:results}

\subsection{Data collection}
\label{subsec:datacollection}

We recruited 176 AMT workers to participate in our conceptualization task. Of these workers, 90 were randomly assigned to the Control group and 86 to the AUI group. These workers completed 1001 tasks: 496 tasks in the control and 505 in the AUI.
All responses were gathered within a single 24-hour period during April, 2017.

After Control and AUI workers were finished responding, we initiated our non-experimental quality ratings task. Whenever multiple workers provided the same response to a given question, we only sought ratings for that single unique question and response.
Each unique question-response pair ($n=428$) was rated at least 8--10 times (a few pairs were rated more often; we retained those extra ratings). 
We recruited 119 AMT workers (who were not members of the Control or AUI groups) who provided 4300 total ratings.

\subsection{Differences in response time}
\label{subsec:differencesInResponseTime}

\begin{figure*}[t]
\centering
{\includegraphics[width=\textwidth]{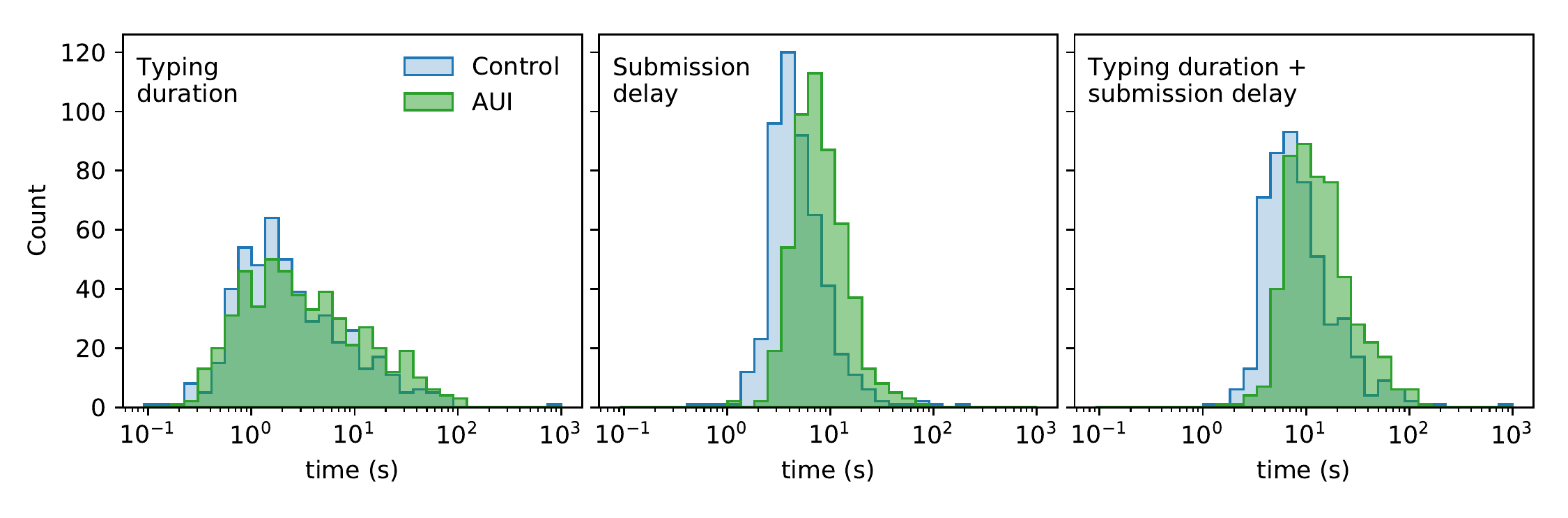}}
\caption{Distributions of time delays. Workers in the AUI treatment were significantly slower than in the control, and this was primarily due to the submission delay between when they finished entering text and when they submitted their response.
\label{fig:distrTimes}}
\end{figure*}

We found that workers were slower overall with the AUI than without the AUI.
In Fig.~\ref{fig:distrTimes} we show the distributions of typing duration and submission delay.
There was a slight difference in typing duration between Control and AUI (median 1.97s for Control compared with median 2.69s for AUI)\footnote{Responses from the AUI group were slightly longer than those from the Control; median length of 11 characters vs.\ 9 characters.}.
However, there was a strong difference in the distributions of submission delay, with AUI workers taking longer to submit than Control workers (median submission delay of 7.27s vs.\ 4.44s). 
This is likely due to the time required to mentally process and select from the AUI options.
We anticipated that the submission delay may be counter-balanced by the time saved entering text, but the total typing duration plus submission delay was still significantly longer for AUI than control (median 7.64s for Control vs.\ 12.14s for AUI).
We conclude that the AUI makes workers significantly slower.

We anticipated that workers may learn over the course of multiple tasks.
For example, the first time a worker sees the AUI will present a very different cognitive load than the 10th time.
This learning may eventually lead to improved response times and so an AUI that may not be useful the first time may lead to performance gains as workers become more experienced.

\begin{figure*}[t]
\centering
{\includegraphics[width=0.475\textwidth]{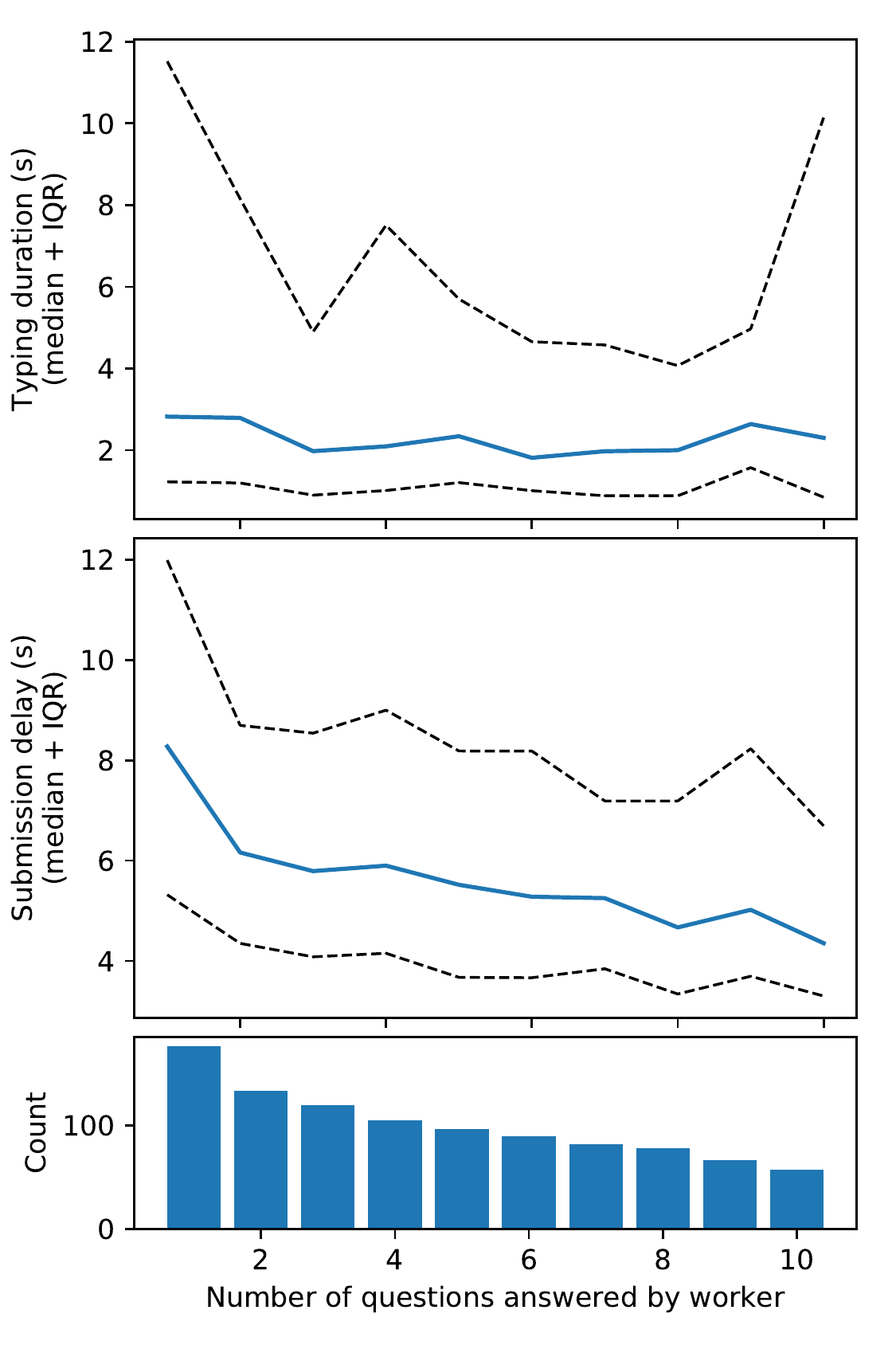}}
{\includegraphics[width=0.475\textwidth]{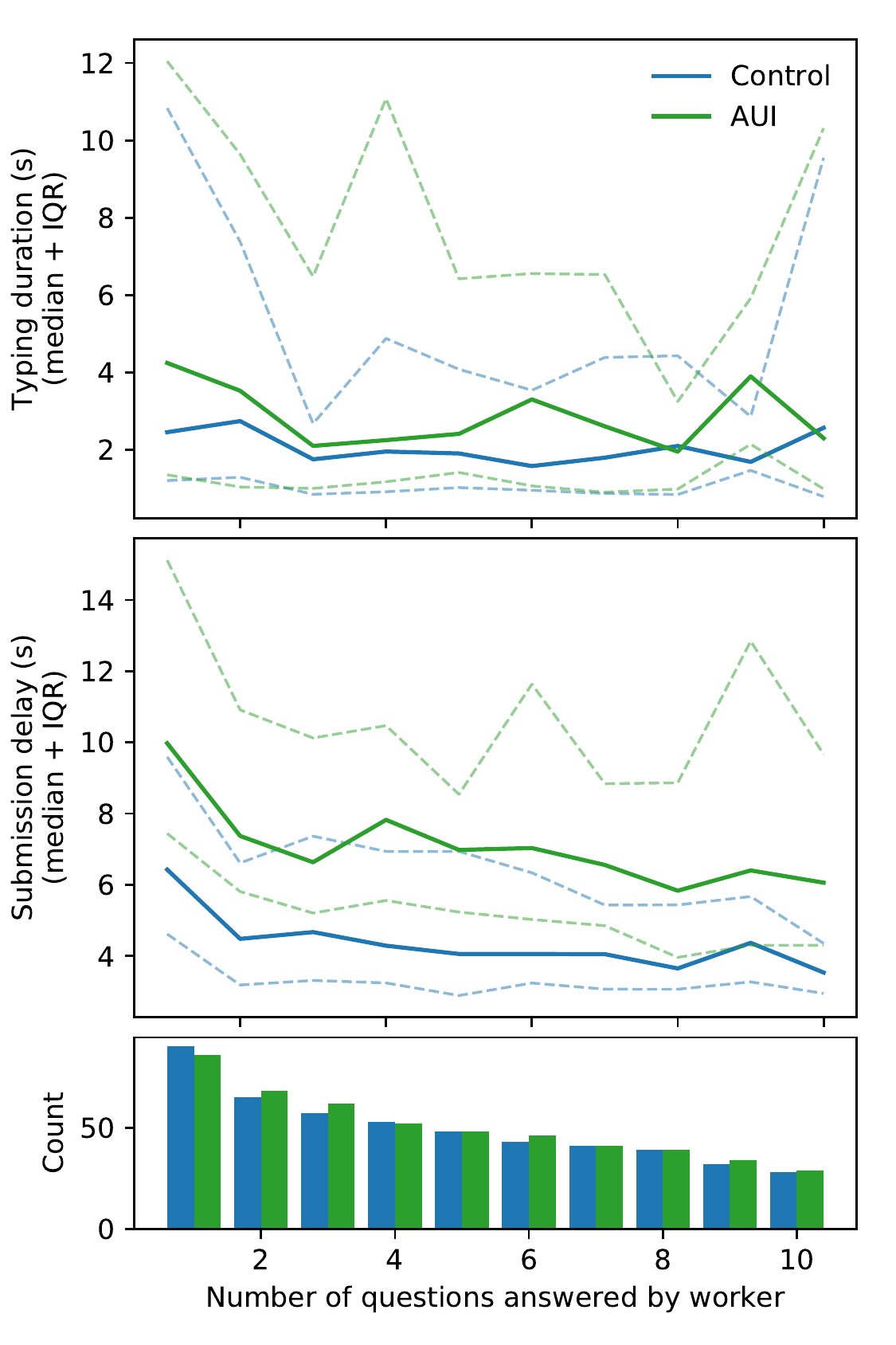}}
\caption{Workers became faster as they gained experience by answering more questions, but
this improvement occurred in both Control and AUI groups.
\label{fig:doWorkersLearn}}
\end{figure*}

To investigate learning effects, we recorded for each worker's question-response pair how many questions that worker had already answered, and examined the distributions of typing duration and submission delay conditioned on the number of previously answered questions (Fig.~\ref{fig:doWorkersLearn}). 
Indeed, learning did occur: the submission delay (but not typing duration) decreased as workers responded to more questions.
However, this did not translate to gains in overall performance between Control and AUI workers as learning occurred for both groups:
Among AUI workers who answered 10 questions, the median submission delay on the 10th question was 8.02s, whereas for Control workers who answered 10 questions, the median delay on the 10th question was only 4.178s.
This difference between Control and AUI submission delays was significant (Mann-Whitney test: $U=872$, $\nA=61$, $\nB=53$, $p < 10^{-4}$).
In comparison, AUI (Control) workers answering their first question had a median submission delay of 10.97s (7.00s).
This difference was also significant (Mann-Whitney test: $U=9822$, $\nA = 169$, $\nB = 165$, $p < 10^{-5}$).
We conclude that experience with the AUI will not eventually lead to faster responses those of the control.

\subsection{Differences in response diversity}
\label{subsec:differencesinresponsediversity}

We were also interested in determining whether or not the worker responses were more consistent or more diverse due to the AUI. 
Response consistency for natural language data is important when a crowdsourcer wishes to pool or aggregate a set of worker responses.
We anticipated that the AUI would lead to greater consistency by, among other effects, decreasing the rates of typos and misspellings.
At the same time, however, the AUI could lead to more diversity due to cognitive priming: seeing suggested responses from the AUI may prompt the worker to revise their response. Increased diversity may be desirable when a crowdsourcer wants to receive as much information as possible from a given task.

\begin{figure*}
\centering
{\includegraphics[width=0.9\textwidth]{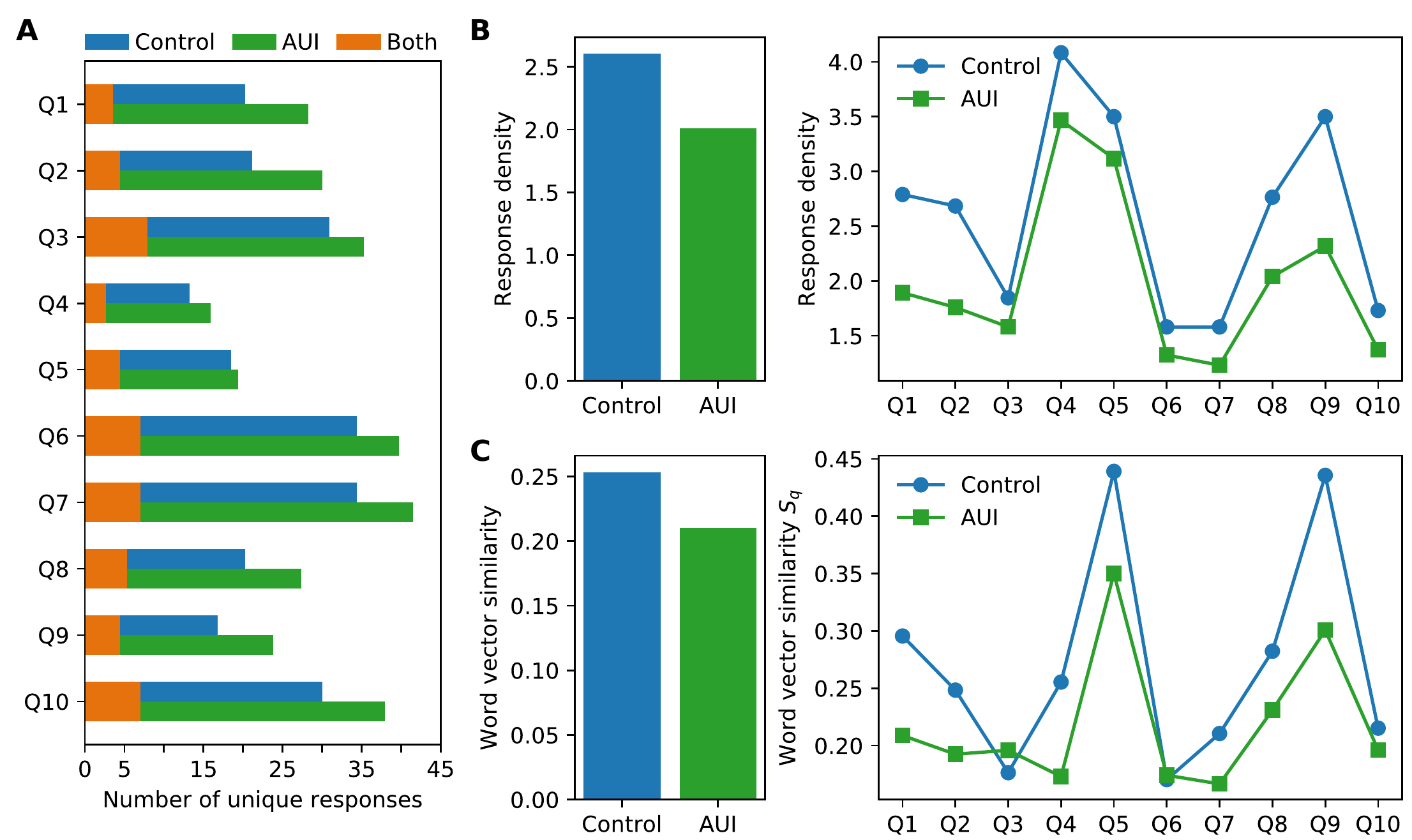}}
\caption{AUI workers had more lexically (A, B) and semantically (C) diverse responses than Control workers.
\label{fig:responseDiversity}}
\end{figure*}

To study the lexical and semantic  diversities of responses, we performed three analyses.
First,
we aggregated all worker responses to a particular question into a single list corresponding to that question. 
Across all questions, we found that the number of unique responses was higher for the AUI than for the Control (Fig.~\ref{fig:responseDiversity}A), implying higher diversity for AUI than for Control.

Second, we compared the diversity of individual responses between Control and AUI for each question. 
To measure diversity for a question, we computed 
the number of responses divided by the number of unique responses to that question. 
We call this the \textit{response density}.
A set of responses has a response density of 1 when every response is unique but when every response is the same, the response density is equal to the number of responses.
Across the ten questions, response density was significantly lower for AUI than for Control (Wilcoxon signed rank test paired on questions: $T=0$, $n=10$, $p < 0.01$) (Fig.~\ref{fig:responseDiversity}B).

Third, we estimated the semantic diversity of responses using word vectors. 
Word vectors, or word embeddings, are a state-of-the-art computational linguistics tool that incorporate the semantic meanings of words and phrases by learning vector representations that are embedded into a high-dimensional vector space~\citep{mikolov2013efficient,NIPS2013_5021}. Vector operations within this space such as addition and subtraction are capable of representing meaning and interrelationships between words~\citep{NIPS2013_5021}.
For example, the vector $\mathbf{v}_\mathrm{king} + \mathbf{v}_\mathrm{woman} - \mathbf{v}_\mathrm{man} $ is very close to the vector $\mathbf{v}_\mathrm{queen}$, indicating that these vectors capture analogy relations.
Here we used 300-dimension word vectors trained on a 100B-word corpus taken from Google News\footnote{\url{https://code.google.com/archive/p/word2vec}} (word2vec). 
For each question we computed the average similarity between words in the responses to that question---a lower similarity implies more semantically diverse answers.
Specifically, for a given question $q$, we concatenated all responses to that question into a single document $D_q$, and averaged the vector similarities $sim(\mathbf{v}_i,\mathbf{v}_j)$ of all pairs of words $(w_i,w_j), w_i \neq w_j$ in $D_q$, where $\mathbf{v}_i$ is the word vector corresponding to word $w_i$:
\begin{equation}
S_q \equiv \frac{\sum_{i=1}^{\left|D_q\right|-1} \sum_{j=i+1}^{\left|D_q\right|} \mathit{sim} (\mathbf{v}_i,\mathbf{v}_j)\left(1-\delta_{ij}\right)}{\sum_{i=1}^{\left|D_q\right|-1} \sum_{j=i+1}^{\left|D_q\right|} \left(1-\delta_{ij}\right)},
\label{eqn:meansim}
\end{equation}
where $\delta_{ij} = 1$ if $w_i = w_j$ and zero otherwise. We also excluded from \ref{eqn:meansim} any word pairs where one or both words were not present in the pre-trained word vectors (approximately 13\% of word pairs).
For similarity $\mathit{sim}(\mathbf{v}_i,\mathbf{v}_j)$ we chose the standard \emph{cosine similarity} between two vectors.
As with response density, we found that most questions had lower word vector similarity $S_q$ (and are thus collectively more semantically diverse) when considering AUI responses as the document $D_q$ than when $D_q$ came from the Control workers (Fig.~\ref{fig:responseDiversity}C).
The difference was significant (Wilcoxon signed rank test paired on questions: $T=4$, $n=10$, $p < 0.05$).

Taken together, we conclude from these three analyses that the AUI increased the diversity of the responses workers gave.

\subsection{No difference in response quality}

\begin{figure*}
\centering
{\includegraphics[width=\textwidth]{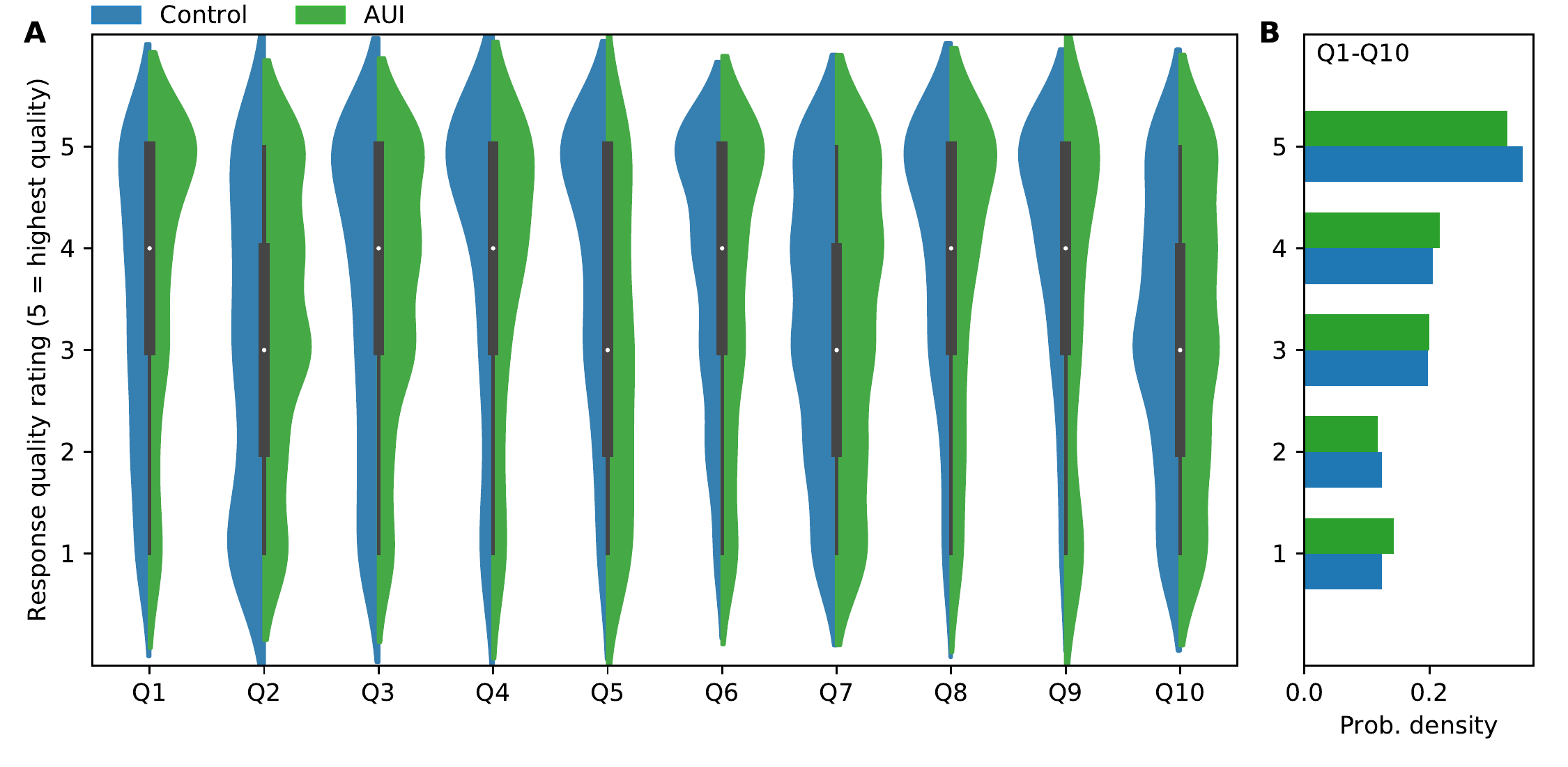}}
\caption{Quality of responses. All question-response pairs were rated independently by workers on a 1-5 scale of perceived quality (1--lowest quality, 5--highest quality).
\label{fig:responseQuality}}
\end{figure*}

Following the collection of responses from the Control and AUI groups, separate AMT workers were asked to rate the quality of the original responses (see Experimental design).
These ratings followed a 1--5 scale from lowest to highest.
We present these ratings in Fig.~\ref{fig:responseQuality}.
While there was variation in overall quality across different questions (Fig.~\ref{fig:responseQuality}A), we did not observe a consistent difference in perceived response quality between the two groups. 
There was also no statistical difference in the overall distributions of ratings per question (Fig.~\ref{fig:responseQuality}B).
We conclude that the AUI neither increased nor decreased response quality.

\section{Discussion}
\label{sec:discussion}

We have showed via a randomized control trial that an autocompletion user interface (AUI) is not helpful in making workers more efficient. Further, the AUI led to a more lexically and semantically diverse set of text responses to a given task than if the AUI was not present. The AUI also had no noticeable impact, positive or negative, on response quality, as independently measured by other workers.

A challenge with text-focused crowdsourcing is aggregation of natural language responses. Unlike binary labeling tasks, for example, normalizing text data can be challenging. Should casing be removed? Should words be stemmed? What to do with punctuation? Should typos be fixed?
One of our goals when testing the effects of the AUI was to see if it helps with this normalization task, so  that crowdsourcers can spend less time aggregating responses.
We found that the AUI would likely not help with this in the sense that the sets of responses became more diverse, not less. 
Yet, this may in fact be desirable---if a crowdsourcer wants as much diverse information from workers as possible, then showing them dynamic AUI suggestions may provide a cognitive priming mechanism to inspire workers to consider responses which otherwise would not have occurred to them.

One potential explanation for the increased submission delay among AUI workers is an excessive number of options presented by the AUI. 
The goal of an AUI is to present the best options at the \emph{top} of the drop down menu (Fig.~\ref{fig:screenshot}B). 
Then a worker can quickly start typing and choose the best option with a single keystroke or mouse click. 
However, if the best option appears farther down the menu, then the worker must commit more time to scan and process the AUI suggestions. 
Our AUI always presented six suggestions, with another six available by scrolling, and our experiment did not vary these numbers.
Yet the size of the AUI and where options land may play significant roles in submission delay, especially if significant numbers of selections come from AUI positions far from the input area. 

\begin{figure}[t]
\centering
\includegraphics[width=0.5\columnwidth]{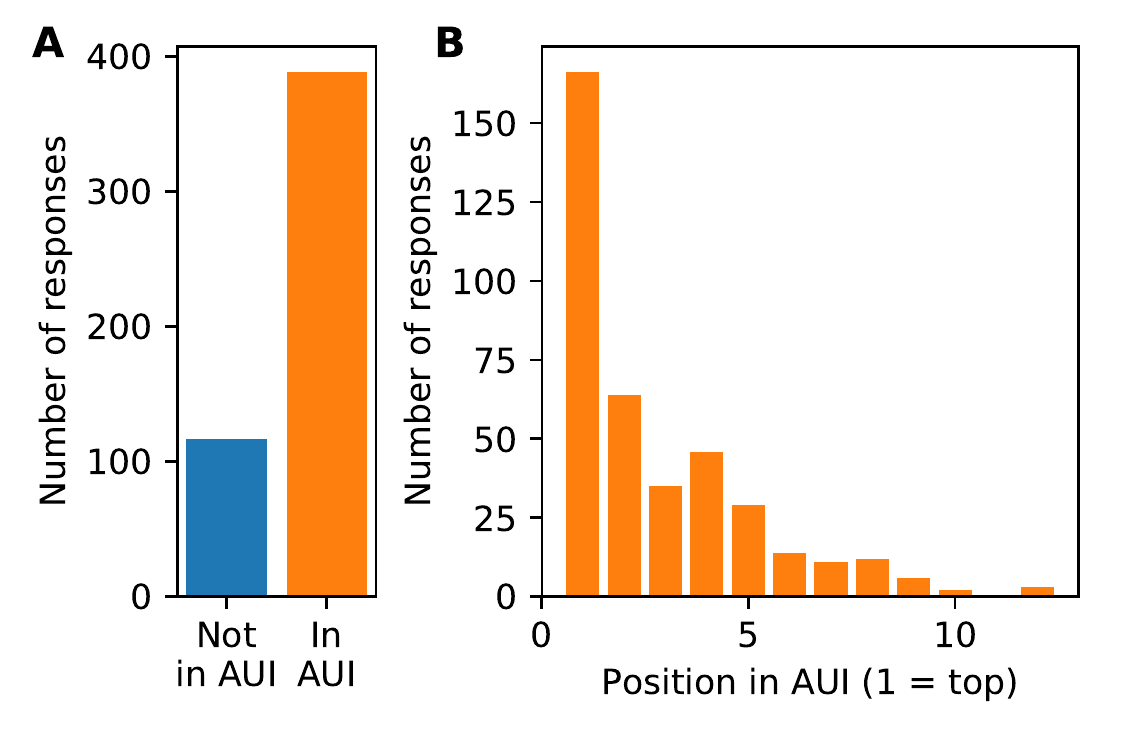}
\caption{Inferred positions of AUI selections based on the last text workers in the AUI group typed before choosing from the AUI.
(\textbf{A}) Most submitted AUI responses appeared in the AUI. 
(\textbf{B}) Among the responses appearing in the AUI, the reconstructed positions of those responses tended to be at the top of the AUI, in the most prominent, accessible area.
\label{fig:AUIpositions}}
\end{figure}

We aimed to explore position effects, but
due to some technical issues we did not record the positions in the AUI that workers chose.  
However, our Javascript instrumentation logged worker keystrokes as they typed so we can approximately \emph{reconstruct} the AUI position of the worker's ultimate response.
To do this, we first identified the logged text inputed by the worker before it was replaced by the AUI selection, then used this text to replicate the database query underlying the AUI, and lastly determined where the worker's final response appeared in the query results.
This procedure is only an approximation because our instrumentation would occasionally fail to log some keystrokes and because a worker could potentially type out the entire response even if it also appeared in the AUI (which the worker may not have even noticed).
Nevertheless, most AUI workers submitted responses that appeared in the AUI (Fig.~\ref{fig:AUIpositions}A) and, of those responses, most owere found in the first few (reconstructed) positions near the top of the AUI (Fig.~\ref{fig:AUIpositions}B). 
Specifically, we found that 59.3\% of responses were found in the first two reconstructed positions, and 91.2\% were in the first six.
With the caveats of this analysis in mind, which we hope to address in future experiments, these results provide some evidence that the AUI responses were meaningful and that the AUI workers were delayed by the AUI even though most chosen responses came from the top area of the AUI which is most quickly accessible to the worker.

Beyond AUI position effects and the number of options shown in the AUI, 
there are many aspects of the interplay between workers and the AUI to be further explored. 
We limited workers to performing no more than ten tasks, but will an AUI eventually lead to efficiency gains beyond that level of experience?
It is also an open question if an AUI will lead to efficiency gains when applying more advanced autocompletion and ranking algorithms than the one we used.
Given that workers were slower with the AUI primarily due to a delay after they finished typing which far exceeded the delays of non-AUI workers, better algorithms may play a significant role in speeding up or, in this case, slowing down workers. 
Either way, our results here indicate that crowdsourcers must be very judicious if they wish to augment workers with autocompletion user interfaces.

\section*{Acknowledgments}

{We thank {S.\ Lehman} and {J.\ Bongard} for useful comments and gratefully acknowledge {the resources provided by the Vermont Advanced Computing Core}. This material is based upon work supported by the National Science Foundation under Grant No.\ {IIS-1447634}.


\end{document}